# Analysis of a power grid using a Kuramoto-like model

**Giovanni Filatrella**
CNR-INFM SuperMat Salerno and Department of Biological and Environmental Sciences, University of Sannio, Via Port'Arsa 11, I-82100, Benevento, Italy

**Arne Hejde Nielsen and Niels Falsig Pedersen**
Oersted-DTU, Section of Electric Power Engineering, The Technical University of Denmark, DK-2800 Lyngby, Denmark

**Abstract:**

We show that there is a link between the Kuramoto paradigm and another system of synchronized oscillators, namely an electrical power distribution grid of generators and consumers. The purpose of this work is to show both the formal analogy and some practical consequences. The mapping can be made quantitative, and under some necessary approximations a class of Kuramoto-like models, those with bimodal distribution of the frequencies, is most appropriate for the power-grid. In fact in the power-grid there are two kinds of oscillators: the "sources" delivering power to the "consumers".

*1.    Introduction*

Synchronization of coupled nonlinear oscillators is a common phenomenon [1] that often can be dealt with using the general framework of the so-called Kuramoto model [2,3]. The model originally proposed by Kuramoto assumes global coupling with each oscillator having its own frequency $\omega_j$ coupled to all other oscillators on an equal footing through the sum of a periodic function and whose strength is determined by a constant K.

$$\dot{\theta}_i = \omega_i + \frac{K}{N}\sum_{j=1}^{N} sin(\theta_j - \theta_i + \delta) \qquad (1)$$

Synchronization is favored if K is large and the frequency spectrum of the oscillators is narrow. With two interacting nonlinear oscillators it has long been known (since the celebrated Huygens experiments on coupled clocks) that they would tend to synchronize [1]. For many oscillators the observation of mutual synchronization is more recent, however examples from the nature are numerous, and are often in qualitative agreement with Eq (1), while some quantitative feature such as the asymptotic behavior for large coupling values require refinements of the model [1,3].

Examples that are cited in review articles on the Kuramoto model are for instance synchronization of the blinks from a group of fireflies [1], synchronization of a number of Josephson junctions coupled to a cavity [4]. People walking on the millennium bridge in London will tend to synchronize their footsteps [5]. The utility power grid has also been suggested [1] as an example of a system of oscillators (power plants producing power at 50 or 60 Hz).

Power plants in a big connected network obviously should be synchronized to the same frequency. If loads are too strong and unevenly distributed or if some major fault or a lightening occurs, an oscillator (power plant) may lose synchronization. In that situation the synchronization landscape may change drastically and a blackout may occur. Examples of such blackouts are numerous, but a well known example is the one in New York in 1965, where about 25 million people were without electricity for up to twelve hours.

Such a catastrophe is well understood in the Kuramoto formalism as a sudden transition from the synchronized state to the incoherent state [3]. We have noticed several references to the continent wide power grid as an example of coupled oscillators [6]; however we have not been able to find in the literature a detailed modeling of the various components in a power grid. In fact on a macroscopic scale a simplified approach to model the dynamics of the grid is necessary, so the network is often simulated without a detailed model for the oscillators [7]. The purpose of this paper is to make physically reasonable models of the various components of a power grid: power plants, transmission lines, distribution lines, transformers, (big) consumers etc.

In the language of electric power engineering a convenient set of units called per units (p.u.) are often used [8]. When these units are used, a high voltage line (e.g. 400 kV) connected to a lower voltage line (e.g. 100 kV) through a transformer and a load at the low voltage side can be considered as a (transformed) load at the high voltage side. This is of course also the case with other unit systems [8] and the complications due to the details of the distribution grid can be avoided by transforming all loads to the same normalized voltage while keeping track of the power flow. We will thus consider all voltages to have the same magnitude across the grid; we also neglect reactive power transfer in the system [8]. A power plant consists of a 'boiler' producing a 'constant' power, as well as a turbine (generator) with high inertia and some damping. In the p.u. units a section of transmission line to a consumer (see Fig 1 ) can be expressed as $P^{MAX}$ sin $\Delta\theta$ (the phase difference between the two ends of the transmission line [8], being $\theta$ a phase angle).

Thus in the equivalent diagram of Fig. 1 the combination of a power plant and a transmission line looks like a power source that feeds energy into a rotating system. This energy can therefore either be accumulated as rotational energy, or be dissipated because of the friction. The remaining part is available for a user (the machine M), provided that there exists a phase angle difference $\Delta\theta$ between the two mechanical rotators that produces the necessary phase shift for ac power transmission. The power flow analysis can therefore be described in terms of the phase angles $\theta$'s that characterize both the rotor dynamics (and hence the energy stored or dissipated) and the power flow between any two rotors connected by an ac line.

With such a model we will then do selected numerical simulations in the spirit of the Kuramoto formalism to establish regions of synchronization as well as regions of instability (blackout) to effectively connect power grid dynamics to the problem of synchronization over a network, a topic of recent great interest [9]. We trust that the application of the method developed to treat Kuramoto-like models can be fruitful for the analysis of power grids. In fact the approaches for coupled oscillators offer an overall view of grid dynamics and provide calculations of the system properties. One might also hope that the Kuramoto approach offers a contribute to an intuitive understanding of the system behavior for a large number of components. Finally, we aim to spur theoretical research towards potentially applicable variants of the Kuramoto model. For this purpose, that we regard as the most relevant, we have developed in detail some particular modifications of the Kuramoto model (containing a bimodal distribution of oscillator frequencies) that make it more appropriate for power grid analysis. The paper is organized in the following way: Section 2 discusses the grid model and sets up the equations to map a power flow analysis on the electrical grid onto a dynamic version of the Kuramoto model. Section 3 presents our numerical simulations for various geographies of the net. The last section contains summary and conclusions.

## 2. *The model.*

We now present the mathematical derivation of the equations in the static case for the stability analysis. In Fig. 1 we show the basic elements of a power grid: an active generator (circle, "G") and a passive machine (square, "M"). The generator converts some source of energy into electrical power, while the reverse is true for a machine. We indicate with $P_{source}$ the power source, or the rate at which energy is fed into the generator. The generator turbine will produce electrical power with a frequency that is close to the standard frequency $\Omega$ of the electric system (50 or 60 Hz, in the equation for the system described below):

$$\theta_1 = \Omega t + \tilde{\theta}_1, \qquad (2)$$

where $\theta_1$ is the phase angle at the output generator. During the rotation the turbine dissipates energy at a rate proportional to the square of the angular velocity ($P_{diss}=Fv \sim v^2$)

$$P_{diss} = K_D \left(\dot{\theta}_1\right)^2, \qquad (3)$$

or it accumulates kinetic energy ½ I $(d\theta_1/dt)^2$ at a rate:

$$P_{acc} = \frac{1}{2} I \frac{d}{dt}\left(\dot{\theta}_1\right)^2, \qquad (4)$$

where I is the moment of inertia. The condition for the power transmission is that the two devices do not operate in phase, and the mismatch between the two rotators is indicated by the symbol "$\Delta\theta$", being the phase difference between the active generator and the passive machine:

$$\Delta\theta = \theta_2 - \theta_1 = \tilde{\theta}_2 - \tilde{\theta}_1. \tag{5}$$

(Let us underline that since all oscillators share the same common frequency $\Omega$ the phase difference between the perturbations (the $\tilde{\theta}$'s) and the original variables (the $\theta$'s) is the same quantity.)

As a function of this phase difference the transmitted power reads:

$$P_{ttransmitted} = -P^{MAX} \sin(\Delta\theta). \tag{6}$$

Thus each generator or machine is described by a power balance equation of the type:

$$P_{source} = P_{diss} + P_{acc} + P_{ttransmitted}. \tag{7}$$

Inserting Eq.s (3-6) into Eq. (7) one gets:

$$P_{source} = I\ddot{\theta}_1\dot{\theta}_1 + K_D(\dot{\theta}_1)^2 - P^{MAX} \sin(\Delta\theta). \tag{8}$$

We assume to be in the limit of small perturbation of the synchronous frequency, see Eq. (2):

$$\dot{\tilde{\theta}}_1 \ll \Omega. \tag{9}$$

Physically, this limit corresponds to the grid operating at a frequency close to the standard 50 or 60 Hz, with very small deviations. Under this hypothesis Eq.(8) can be approximated with:

$$P_{source} \cong I\Omega\ddot{\tilde{\theta}}_1 + \left[ I\ddot{\tilde{\theta}}_1 + 2K_D\Omega \right]\dot{\tilde{\theta}}_1 + K_D\Omega^2 - P^{MAX} \sin(\Delta\theta). \tag{10}$$

We will now assume that the coefficient of the first derivative is constant, neglecting the term proportional to the acceleration;

$$\ddot{\tilde{\theta}}_1 \ll 2K_D\Omega/I. \tag{11}$$

In practical terms we assume that the rate at which energy is stored in the kinetic term ($\approx I\Omega\ddot{\tilde{\theta}}_1$) is much less of the rate at which energy is dissipated by friction ($\approx K_D\dot{\tilde{\theta}}_1^2 \cong K_D\Omega^2$), a common condition for mechanical systems. Under this hypothesis Eq. (10) becomes:

$$I\Omega\ddot{\tilde{\theta}}_I = P_{source} - K_D\Omega^2 - 2K_D\Omega\dot{\tilde{\theta}}_I + P^{MAX}\sin(\Delta\theta). \quad (12)$$

Such equation can be cast into normalized units:

$$\ddot{\tilde{\theta}}_I = P - \alpha\dot{\tilde{\theta}}_I + P^{MAX}\sin(\Delta\theta), \quad (13)$$

that is formally identical to the basic ingredient of the Kuramoto model [2,3].

So far we have considered the deviations from the predetermined frequency $\Omega$ of the grid to compute the stability analysis of the deviations from the uniform rotation. Knowing a priori the rotating frequency one could subtract the uniform rotation [Eq. (2)], thus reducing the problem to a static solution. However, the mapping onto some Kuramoto model can be exploited to a much larger extent to predict the dynamic stability of the system *without* the assumption of a predetermined frequency. In fact the Kuramoto model has been proposed to solve the much more complicated problem of mutual synchronization of different oscillators when the synchronization in frequency is not guaranteed. So we now turn to the problem of N oscillators of the type Eq. (12). Let us assume that even subtracting the common frequency $\Omega$ each oscillator will have a natural average velocity close to $\Omega$, the deviation being denoted by the frequency $\omega_i$:

$$\langle\dot{\theta}_i\rangle = \Omega + \omega_i. \quad (14)$$

The $\omega_i$ are the spreads (due to the differences in the oscillators parameters) around zero; spreads that are small, $\omega_i \ll \Omega$. This corresponds to the fact that power generators can actually supply ac voltages with frequency slightly different from the standard reference, see Fig. 2. The origin of the frequency shift in model Eq. (12) stems from the presence of different power sources that are moreover spread around two different peaks: a positive input power for generators and a negative power peak for machines. So we should re-write the model Eq. (12) in the form:

$$\ddot{\tilde{\theta}}_i = \left[\frac{P^i_{source}}{I\Omega} - \frac{K_D\Omega}{I}\right] - \frac{2K_D}{I}\dot{\tilde{\theta}}_i + \frac{P^{MAX}}{I\Omega}\sin(\Delta\theta), \quad (15)$$

where the index i for the power source recalls that the input power depends upon the site. For completeness let us add that we are assuming that dissipation is the same for all sources. Eq. (15) reduces to a Kuramoto-like model:

$$\ddot{\tilde{\theta}}_i = w_i - \alpha \dot{\tilde{\theta}}_i + K \sum_{j \neq i} a_{j,i} \sin(\tilde{\theta}_j - \tilde{\theta}_i) \qquad (16)$$

with new renormalized parameters $w_i$, $\alpha$, $K$, and time to be defined below. The adjacency matrix $a_{i,j}$ accounts for the topology of the grid [7], it reads 1 (or any other finite value) if the nodes i and j are connected, 0 otherwise. In the all to all coupled model $a_{i,j}=1$ $\forall$ i,j. When time is normalized with respect to a common frequency $\Omega^{-1}$, the parameter $\alpha$ for the dissipation reads:

$$\alpha = \frac{2K_D \Omega}{I}. \qquad (17)$$

The coupling constant K is:

$$K = \frac{P^{MAX} \Omega}{I}. \qquad (18)$$

The $w_i$'s are not centered around zero:

$$w_i = \left( \frac{P^i_{source} \Omega}{I} - \frac{K_D \Omega^3}{I} \right) = W_i + w_0. \qquad (19)$$

Not surprising we have obtained the same constant shift of the frequencies as in Eq. (12) plus a distribution of the frequencies. A further change of variables to move in the rotating frame $w_0 t$ leads, without loss of generality, to the Kuramoto-like model:

$$\ddot{\tilde{\theta}}_i = W_i - \alpha \dot{\tilde{\theta}}_i + K \sum_{j \neq i} a_{j,i} \sin(\tilde{\theta}_j - \tilde{\theta}_i). \qquad (20)$$

As a final remark on the distribution of the $W_i$ let us underline that the power source can be either positive (generators) or negative (machines) Therefore one expect a bimodal distribution of the frequencies $W_i$. Nevertheless some important results have been derived for the Kuramoto model in presence of bimodal distributions for the case of global coupling ($a_{i,j}=1$). A number of studies have been devoted to the massless Kuramoto model in the presence of a bimodal distribution of the oscillators [10], also in presence of a time delay [11]. The underdamped case has been considered in Ref. [12]. We can therefore straightforwardly predict that in the case of all-to-all (global) coupling the synchronization will occur for large enough coupling K. Moreover, for the underdamped Kuramoto model it has been predicted that the synchronization will be hysteretic [13]: increasing the coupling from the non synchronized solution will not lead to phase-lock until a critical value which is higher than the corresponding value obtained decreasing the coupling from the synchronized mode. Processes such as the insertion of oscillators one by one has been derived in the case of unimodal compact distributions [14] and with a modification of the Kuramoto model for a generic distribution [15]. In both cases a sudden transition to the coherent state is predicted. However the relevant case here, Eqs. (19,20) of a bimodal distribution of the oscillators on a network with local connections has not yet been considered, to our knowledge.

## 3. Numerical results

The combination of equations (13) amounts circuitwise to the combination of elements shown in the diagram of Fig. 1. So generators and machines can be assembled to form a circuit and the corresponding equation can be derived. Few examples are shown in Fig. 3, ranging from the simplest Generator-Machine system (3a) to an approximation to the Zealand (Denmark) power grid system (3c). When the parameters of the "circuit" elements are given the corresponding equations are promptly derived, so the network in Fig. 3a is described by the normalized equations (the tilde is omitted for brevity):

$$\ddot{\theta}_1 = -\alpha\dot{\theta}_1 + P_1 + P^{MAX}\left[\sin(\theta_2 - \theta_1) + \sin(\theta_3 - \theta_1)\right]$$
$$\ddot{\theta}_2 = -\alpha\dot{\theta}_2 + P_2 + P^{MAX}\left[\sin(\theta_1 - \theta_2)\right] \quad (21)$$
$$\ddot{\theta}_3 = -\alpha\dot{\theta}_3 + P_3 + P^{MAX}\left[\sin(\theta_1 - \theta_3)\right]$$

Here for simplicity all transmission lines have been assumed identical, and therefore described by the same parameter $P^{MAX}$. The same holds for the dissipation parameter $\alpha$. The parameter $P_i$ contains the information on the nature of the device: it is positive for a generator that is a source of power, and negative for an absorbing machine.

The equations for Fig. 3c are notationally more involved, but straightforward. For instance node 1 corresponds to a generator governed by the equation:

$$\ddot{\theta}_1 = -\alpha\dot{\theta}_1 + P_1 + P_{1A}^{MAX}[\sin(\theta_A - \theta_1)] + P_{1C}^{MAX}[\sin(\theta_C - \theta_1)] \qquad (22)$$

and, for the particular case here examined, $P_1=2.5$, $P_{1A}^{MAX}=7$, $P_{1C}^{MAX}=10$.

Next we explore the behavior of the simplest layout (3a) to illustrate the network analysis we can perform with this simplified reduction. Consider the connection of Fig. 3a, with a machine absorbing 2 power units and two generators delivering 1 power unit each. We analyze the behavior of the system under a sudden short perturbation of the type depicted in Fig. 4a: the machine for a short while requires more power than under standard operation (3 units instead of 2); this extra energy is taken from the kinetic energy of the rotators that after few time units restores normal operation. This type of perturbation is not the standard stability analysis; rather it constitutes a realistic model of an unbalanced power due to a short circuit fault. Depending on the system parameters two outcomes are possible:

1) the network after the perturbation returns to stable operation (Fig. 4b) after a perturbation $\Delta P=1$ for a duration $\Delta t=2$, in spite of the oscillations in the delivered power (solid line). The new stable point is phase shifted with respect to the pre-crisis point of an integer multiple of $2\pi$ (dashed line). In this simulation the return to the synchronous motion is achieved without a change of the external parameter by the sole stability property of the system;

2) the network after the perturbation does not return to stable operation (Fig. 4c) after a perturbation $\Delta P=1.5$ for a duration $\Delta t=2$, because the oscillations of the delivered power (solid line) stay. The phase continuously shifts with some average slope (notice the break of the axis) of about 24 rad/time unit.

We conclude that in the case 1) the system is capable to withstand the perturbation, while in the case 2) the network would lose the stability even after the end of the perturbation. With this definition in mind we have analyzed a few cases where intuition and simulations of the most complete model give a reference for the expected behavior. We have begun with system 3a and varied the time duration of the perturbation, the basis of the peak in Fig. 4a. It is expected [8] that the relevant parameter is the total energy of the disturbance, so we have simulated several pulses with different duration and we have sought for the minimum height that produces a permanent instability of the type shown in Fig. 4b. The results are shown in Fig. 5a. Two features are evident:

a) for short pulses ($\Delta t < 0.5$) the behaviour is nonlinear in the energy.
b) For long pulses ($\Delta t > 2$) there is enough time for the system to reach an equilibrium with the perturbation, so prolonging the perturbation has no effect on the instability region.

Both results are in qualitative agreement with the network analysis [8], thus confirming that even this very simplified version of the power grid can capture some relevant features of the real system

Next we have analyzed the effect of the topology of the connection on a simple network, see Fig. 3b. The question we have asked is: will the system be capable to tolerate a more severe shock when a third generator (with respect to 3a) is just connected to the other two generators (we call this A configuration), or when also a direct connection to the absorber is inserted (we call this B configuration, see Fig. 3b with dashed line active) or finally when the two generators nearest to the machine are directly connected (C configuration, fig. 3b with the dotted line active). To avoid some special value of the parameters we have systematically investigated a range for the transmission line capability (x-axis, Fig. 5b). For reference we also show the result for the two-generator case. The figure demonstrates that the direct connection improves the stability of the system even if the absorbed power is increased (having 3 generators we have set the absorbing power of the machine to 3 units), while a direct connection between the two satellite generators does not improve the stability properties.

Also in this case the purpose is to confirm that the Kuramoto-like model can reproduce at least qualitatively the more accurate network analysis [8].

In Fig. 6 we have analyzed the reaction of the network to disturbances occurring in different points of the system. The vertical axis reports the minimum value for the disturbance ΔP to induce a switch to the unstable operation mode, similar to the transition observed in Fig. 4c. As for the previous case we have investigated several values for the power transmission parameters to detect systematic trends. In fact, the behavior is clearly consistent with the general expectation that stronger connections tend to stabilize the system: the larger the value of the connection $P^{max}$, the larger the perturbation required to drive the system out of equilibrium. We also notice that independently of the machine where the perturbation occurs, the three curves have approximately the same slope.

In Fig. 7 we have tried to elucidate the effect of inertia on the stability of the system. Considering that the consumers are not just motors but often purely resistive devices it is reasonable to assume that the dissipation will be higher in the machines (consumers) than in the generators. We have therefore increased the dissipation α for the consumers in the network 3c and found the minimum value of a disturbance capable to destabilize the system, analogous to the threshold found in Fig. 6. In general the system becomes more unstable increasing the dissipation of the consumers, but the trend is not obvious and sometimes even shows windows of poorer stability. This proves that even for moderately complicated network the response to strong perturbation can be rather involved.

*Conclusions*

A utility power grid has been analyzed to get insight into the complicated dynamics that determines the behavior of the power grid. We found that the dynamics can be described in the framework of the Kuramoto theory for coupled oscillators.

Beside the formal analogy, that has its interest *per se*, it would be of course also interesting to apply the body of knowledge accumulated for the Kuramoto model to solve practical problems of the utility power grid. Let us use here a word of caution. The approximations employed to derive the analogy are to some extent relevant, for instance we completely neglect the so called active control that continuously monitors power distributions and tries to maintain the system in a stable state. For a simple system of few generators and users it makes little sense to use the approximated equations, when the usual engineering methods are much more accurate. However, the power grid is becoming increasingly articulated, and the use of many small power sources (referred to as microgenerators, i.e. small windmill or photovoltaic roofs on private homes), that are not controlled by a centralized management is increasing. It is in this direction that we believe a simple physical model can help, suggesting *qualitative* design rules of the network that favor stable operation. However, as we have mentioned, the appropriate local Kuramoto model with inertia and bimodal distribution is still unexplored – not to speak of the necessity to extract from the predictions results that are robust enough to be applied to a practical grid. We hope this paper can anyway constitute a stimulus for research in such direction.

GF wishes to thank the financial support from the ESF network-programme "Arrays of Quantum Dots and Josephson Junctions". NFP was supported by the Danish STVF program "New Superconductors".

# Power

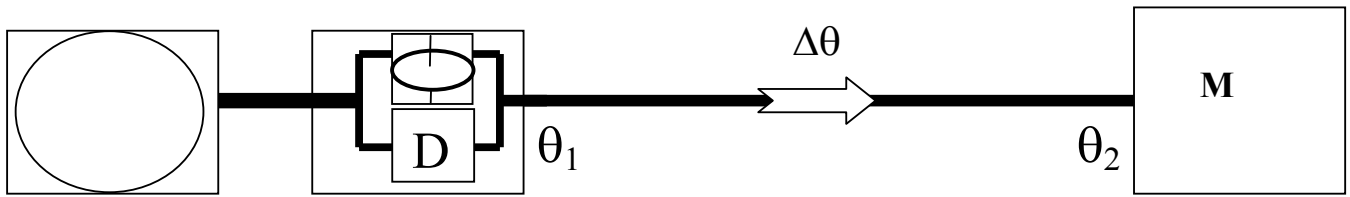

**FIG. 1: Equivalent diagram of generator and machine connected by a transmission line. The turbine consists of a flywheel and dissipation D.**

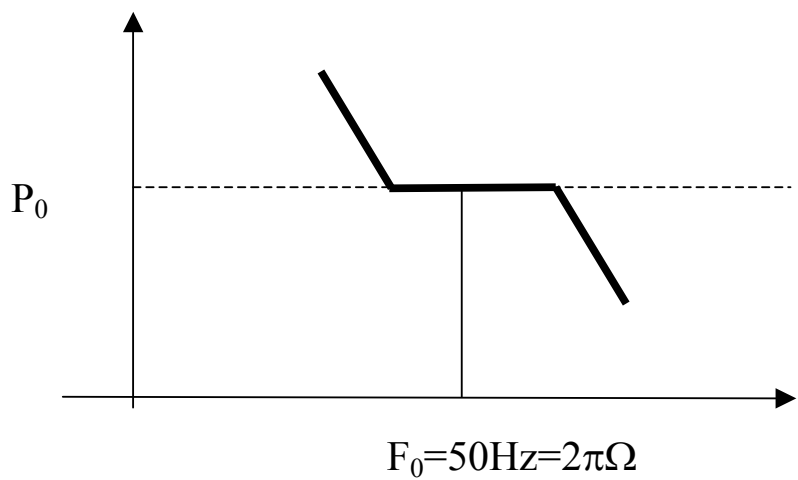

Fig. 2: Power-frequency characteristic of electrical plant.

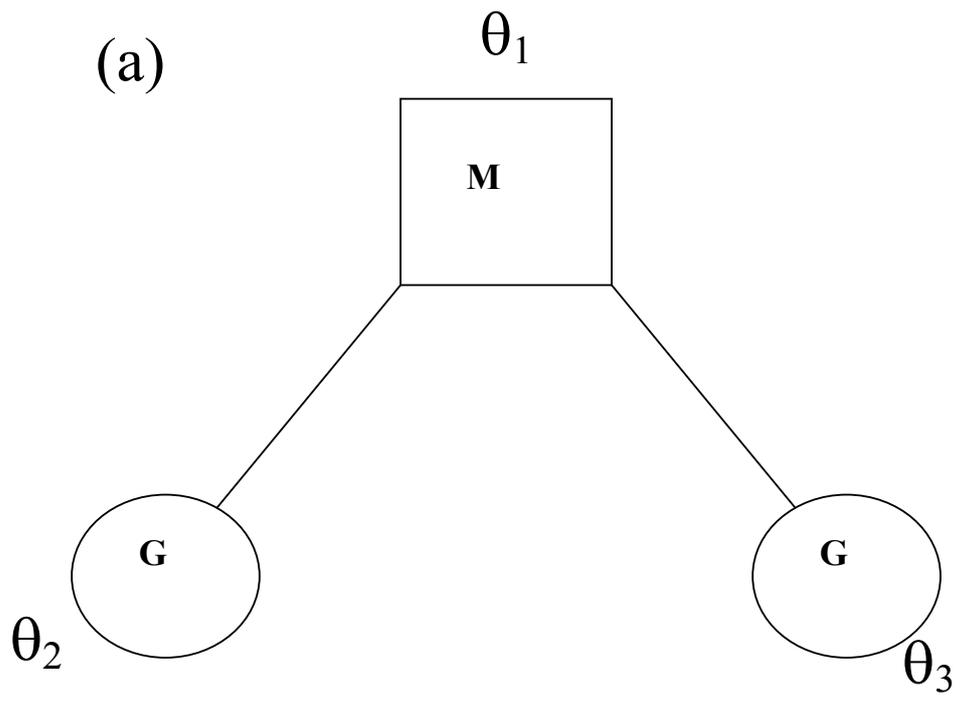

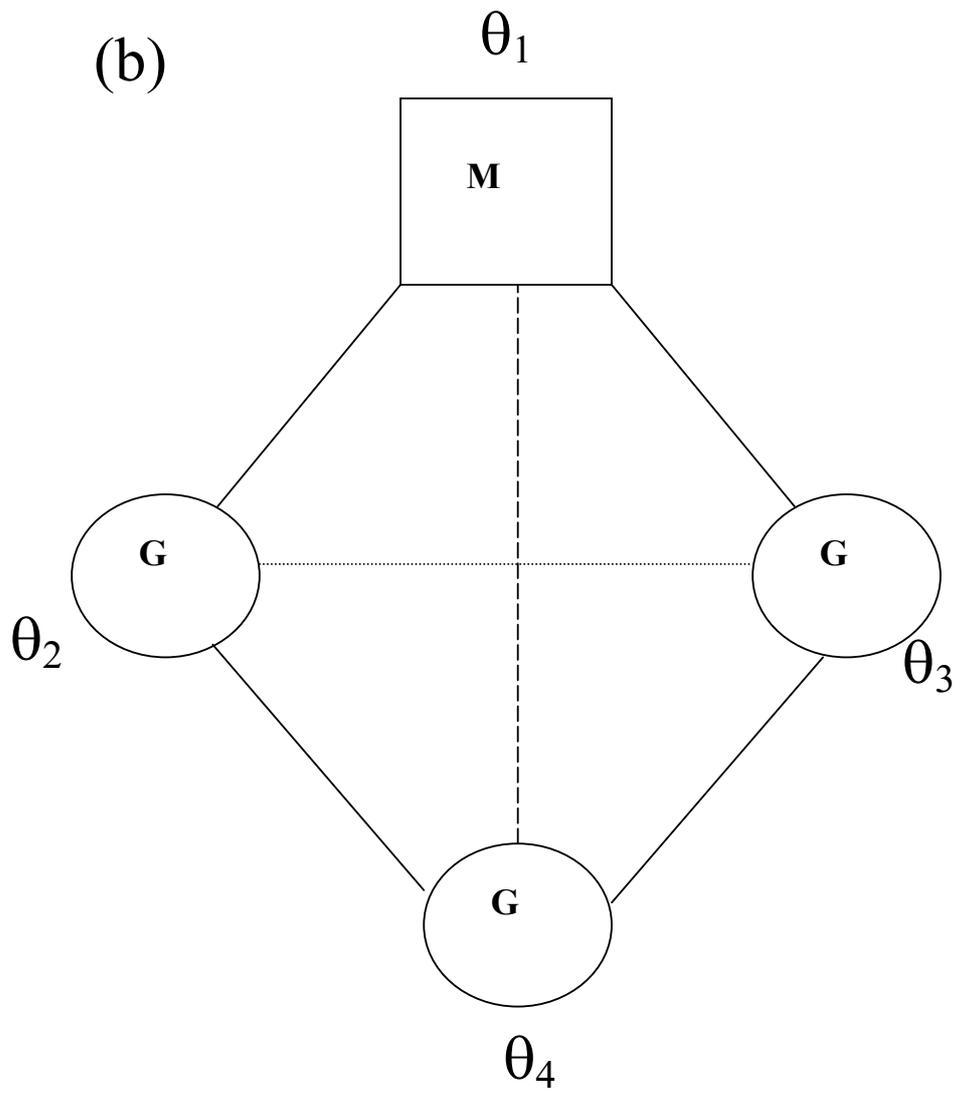

(c)

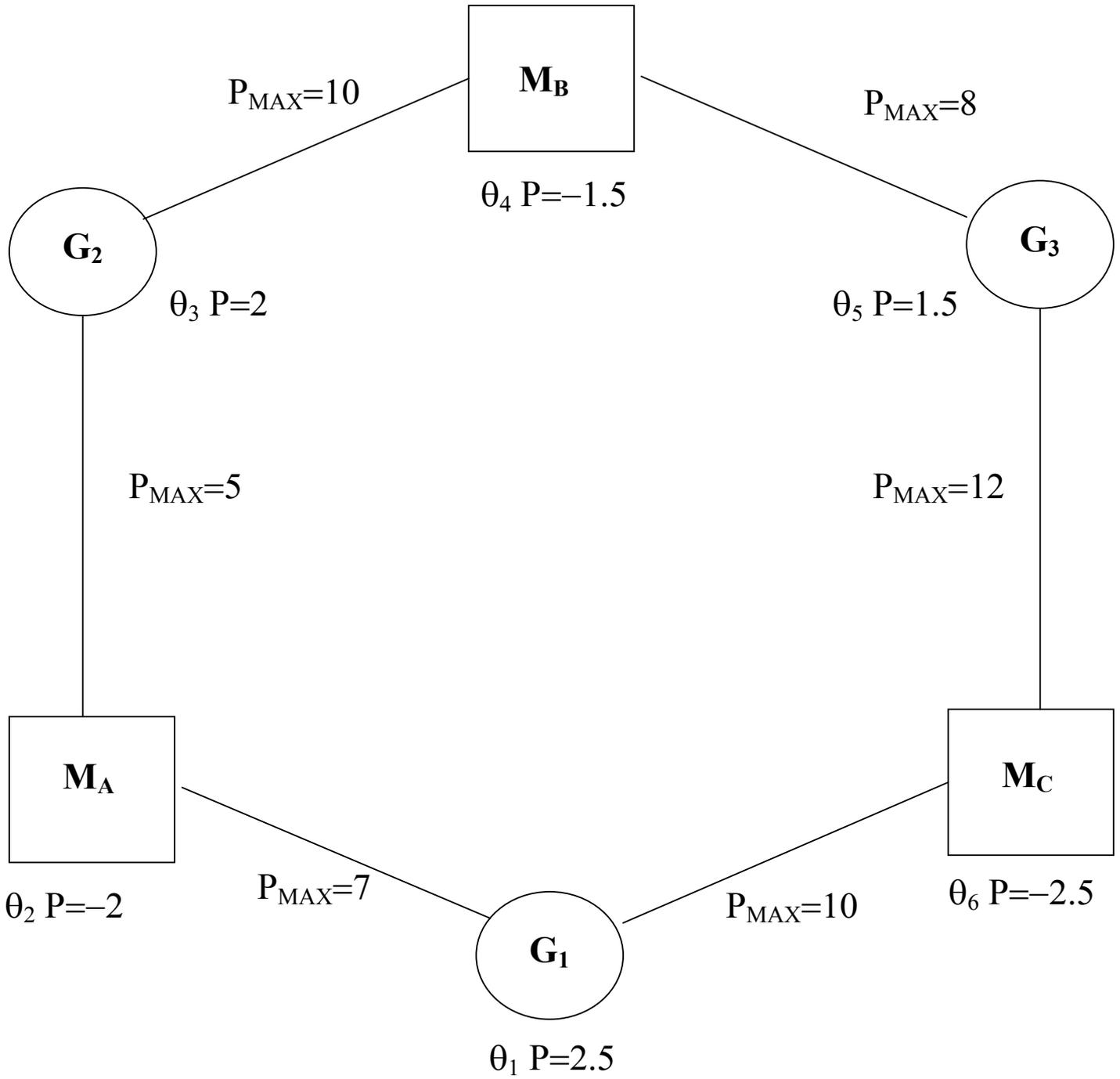

**Fig. 3:**

a) Connection of two generators with a machine.

b) Connection with three generators with a machine (model "A" connection in fig. 5b). The dotted line represents an additional cross connection between two generators (model "B" connection in fig. 5b), the dashed line an additional direct link of the machine with the remote generator (model "C" connection in fig. 5b).

c) Connections of three generators and three machines as in an approximation to the Zealand (Denmark) power grid system.

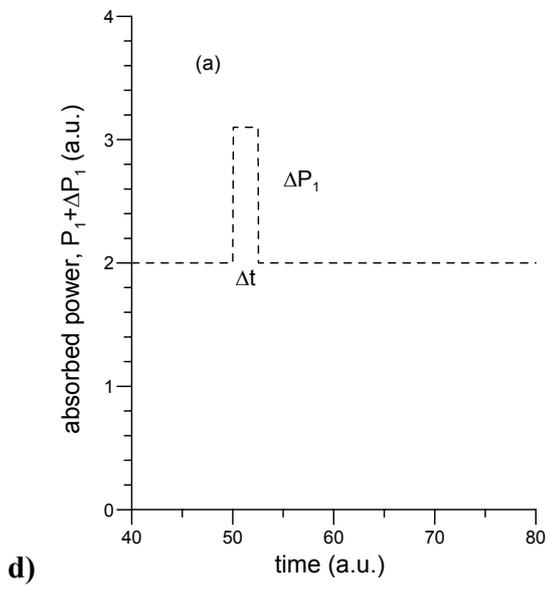

**d)**

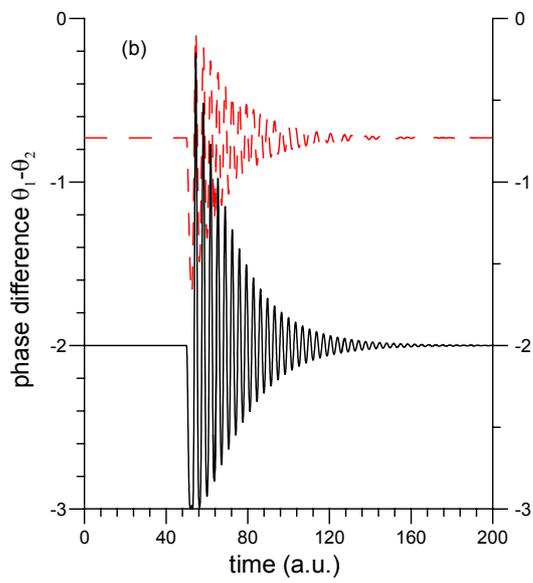
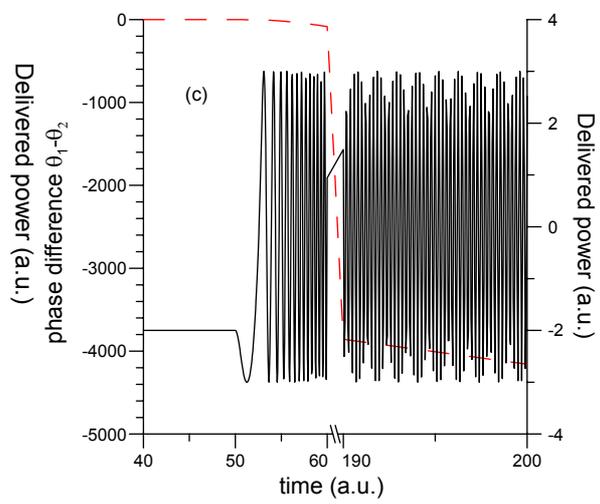

**FIG. 4: Circuit of Fig. 3a, phase dynamics of the machine and power delivered to the machine. Each generator produces one normalized unit of energy, while the machine consumes two units. Dissipation is $\alpha=0.1$.**

**a)** Time dependence of the perturbation of the power absorbed by the machine.

**b)** Stable operation after a perturbation. The phase difference is denoted by a dashed line (left axis) while the power delivered to the machine by a solid line (right axis). The perturbation (height $\Delta P$ of Fig 4a) is 1 unit. The duration is $\Delta t = 2$ units.

**c)** Unstable operation after a perturbation. The notation is the same as in Fig. 4b. The perturbation (height $\Delta P$ of Fig 4a) is 1.5 unit. The duration is $\Delta t = 2$ units.

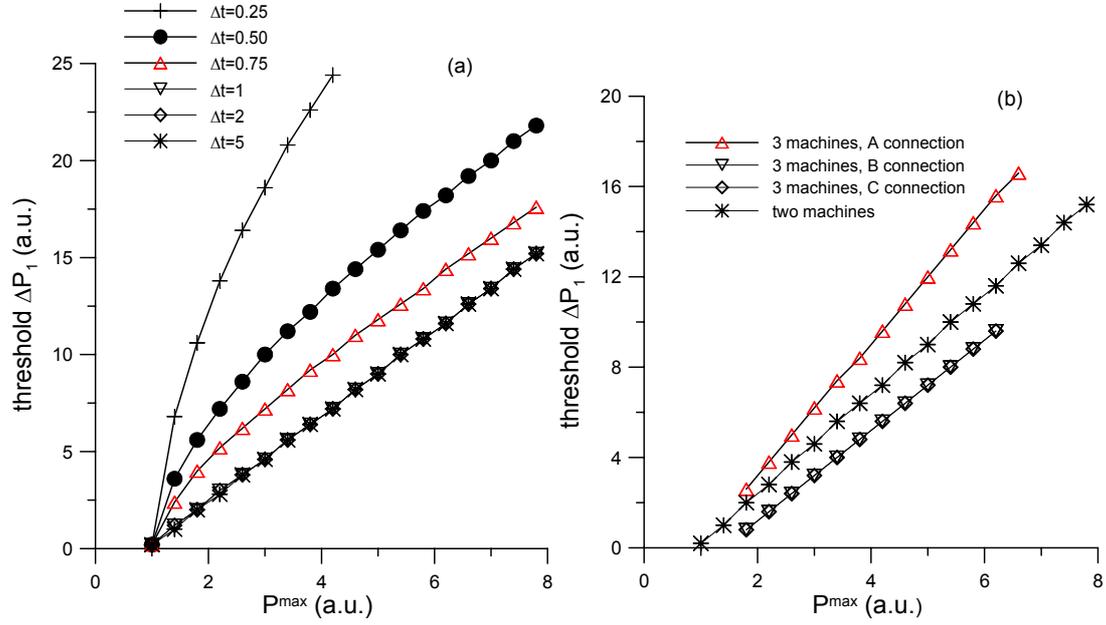

**FIG. 5: Analysis of the stability of the system against perturbations. The threshold is the minimum perturbation (see Fig. 4c) that causes the transition from stable (Fig. 4a) to unstable (Fig. 4b) operation.**

**a) Behavior as a function of the transmission lines capabilities (PMAX) for the system of Fig. 3a varying the perturbation duration, see Fig. 4c.**

**b) Behavior as a function of the transmission lines capabilities (PMAX) for the systems of Fig. 3a and 3b varying the connections, see caption Fig. 3b. Each generator produces one normalized unit of energy, while the machine consumes three units. Dissipation $\alpha$ is 0.1.**

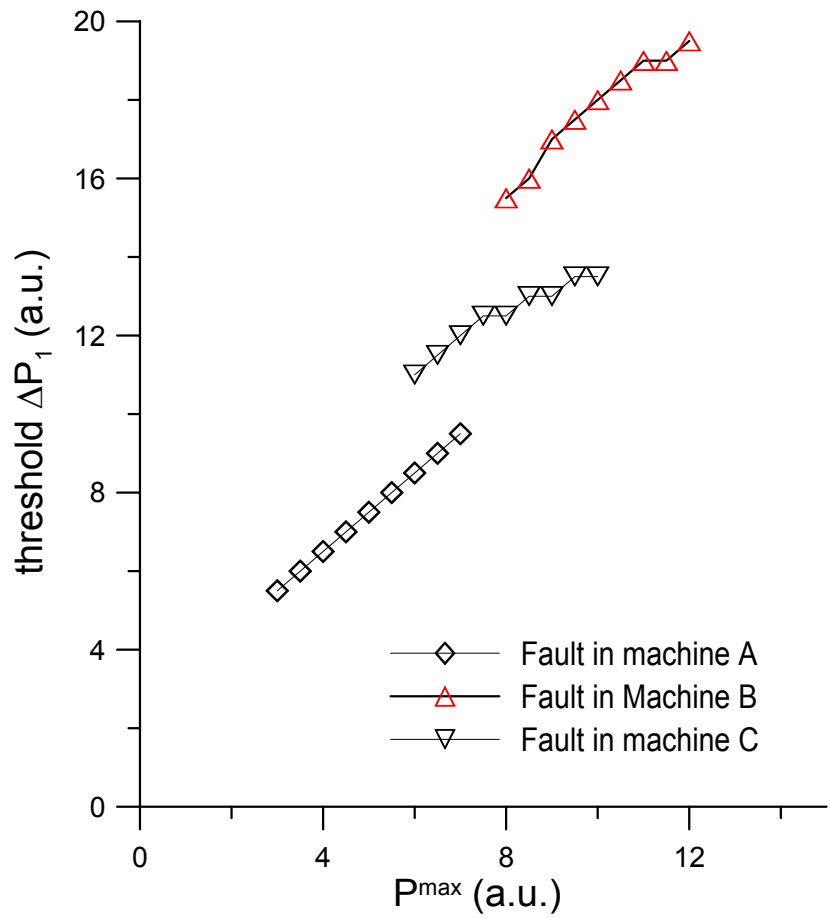

**FIG. 6. Stability of the nine bus system as a function of the perturbation in the various consumers for different values of the transmission power capability. Parameters are the same as in Fig. 3c. Dissipation is set to 0.1.**

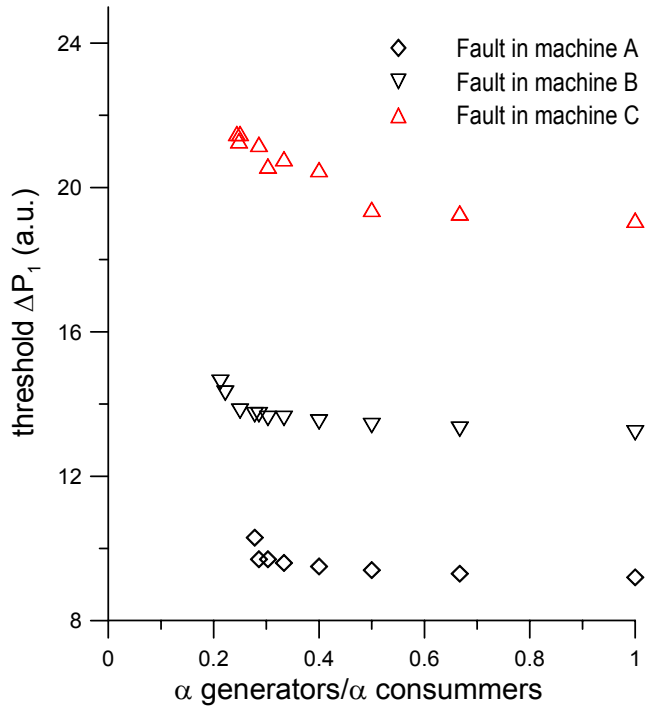

**Fig. 7 Stability of the nine-bus system as in Fig. 3c varying the dissipation of the machines and perturbing one machine at a time. The duration of the perturbation is Δt = 2 units.**